\documentclass{iau}
\usepackage{graphicx}

\title[IAUS291.~~DM variation \& MSP timing] 
{Tracking dispersion measure variations of timing array pulsars with the GMRT} 

\author[U. Kumar et al.] 
{Ujjwal Kumar$^1$, Yashwant Gupta$^1$, Willem van Straten$^2$, \\ Stefan Os{\l}owski$^{2,4}$, Jayanta Roy$^1$, N. D. R. Bhat$^{2,3}$, Matthew Bailes$^2$ \and Michael J. Keith$^4$}

\affiliation{$^1$National Centre for Radio Astrophysics, Pune University Campus, Postbag 3, India 411007 \\ email: {\tt ujjwal@ncra.tifr.res.in} \\[\affilskip]
$^2$Centre for Astrophysics \& Supercomputing, Swinburne University of Technology, Australia \\[\affilskip]
$^3$International Centre for Radio Astronomy Research, Curtin University, Australia \\[\affilskip]
$^4$Australia Telescope National Facility, CSIRO, Australia}

\pubyear{2012}
\volume{291}  
\jname{\mbox{Neutron Stars and Pulsars: Challenges and Opportunities after 80 years}}
\editors{J. van Leeuwen, ed.} 
\begin{document}

\maketitle

\begin{abstract}
We present the results from nearly three years of monitoring of the variations in dispersion measure (DM) along the line-of-sight to
11 millisecond pulsars using the Giant Metrewave Radio Telescope (GMRT). These results demonstrate accuracies of single epoch DM 
estimates of the order of $5 \times 10^{-4} \rm{~cm^{-3}~pc}$. A preliminary comparison with the Parkes Pulsar Timing Array (PPTA) data shows
that the measured DM fluctuations are comparable. We show effects of DM variations due to the solar wind and solar corona and compare with the existing models.
\end{abstract}


\firstsection 
\section{Introduction}
Dispersion measure quantifies the integrated dispersive effect of the plasma between the pulsar and the observing telescope, on the propagating broadband pulsar signal.
In general, it varies with time due to reasons such as the transverse
motion of pulsar sampling different lines of sight (LOS) through inhomogeneous and turbulent 
interstellar medium (ISM), solar wind and solar corona, plasma density changes in the binary orbit and drifting wisps of ionized gas in supernova shell.
For Pulsar Timing Arrays (PTAs), which aim for a final accuracy of 100 $\rm{ns}$ or better at L band, DM variations as small as $\sim 5 \times 10^{-5} 
\rm{~cm^{-3}~pc}$ need to be corrected for. Meanwhile, the timing accuracies currently achieved for most of the PPTA pulsars are still of the order of 
a $\rm\mu s$ and above (\cite{Manch11}) and therefore DM corrections could improve these.
As indicated from the observations by \cite{Back93}, \cite{Hobbs04},
later from analytical derivation, for a turbulent ISM, 
$|d(DM)/dt| \approx 0.0002 ~\sqrt{DM} \rm{~cm^{-3}~pc~yr^{-1}}$, which implies significant change over a period of a few days to a week for a typical DM 
of a few tens of $\rm{cm^{-3}~pc}$.
The GMRT, using its low frequency capability, can provide more accurate DM measurements by taking advantage of the inverse-square 
law dependency of the delay on the observing frequency, as has been demonstrated by \cite{Ahuja05}, who had achieved an accuracy of up to 
$5 \times 10^{-3} \rm{~cm^{-3}~pc}$ for long period pulsars.

\section{Observations and analysis}
A program was initiated at the GMRT, in Nov 2009, to carry out roughly
bi-weekly simultaneous dual-frequency observations at 325 and 610 MHz
for 11 millisecond pulsars (MSPs),
primarily to track the DM variations accurately and study their effects on timing accuracy as well as for studying DM variations due to the solar corona 
and the solar wind. The observations used the GMRT software back-end (\cite{JRoy10}) in the simultaneous dual-frequency phased array mode, giving total intensity time-series 
from 512 channels over 32 MHz of bandwidth at each frequency. In this mode the data streams from the two frequencies are locked to each other without any instrumental 
delay, allowing accurate DM estimates without requiring absolute timing measurements.
The data were incoherently dedispersed and folded using a Doppler corrected period. The delay was computed using the peak of the cross-correlation between
the profiles at the two frequencies. The DM was computed as $DM =  (\Delta{t}/K) \times 1/(\nu_1^{-2}-\nu_2^{-2}) \rm{~cm^{-3}~pc}$,
where K, called dispersion constant, is equal to $4.1488080 ~(\pm30) \times 10^3 \rm{~MHz^2~cm^3~pc^{-1}~s}$ and $\Delta t$ is the total delay, as seen 
at the solar system barycenter, between the signals at the two
frequencies $\nu_1$ and $\nu_2$. Errors were estimated by propagating the off-pulse noise of the two profiles to estimate the rms error of the measured delay.

\section{Results, conclusions \& future goals}
\begin{figure*}[tb]
\begin{center}
\includegraphics[height=2.0in,width=2.7in]{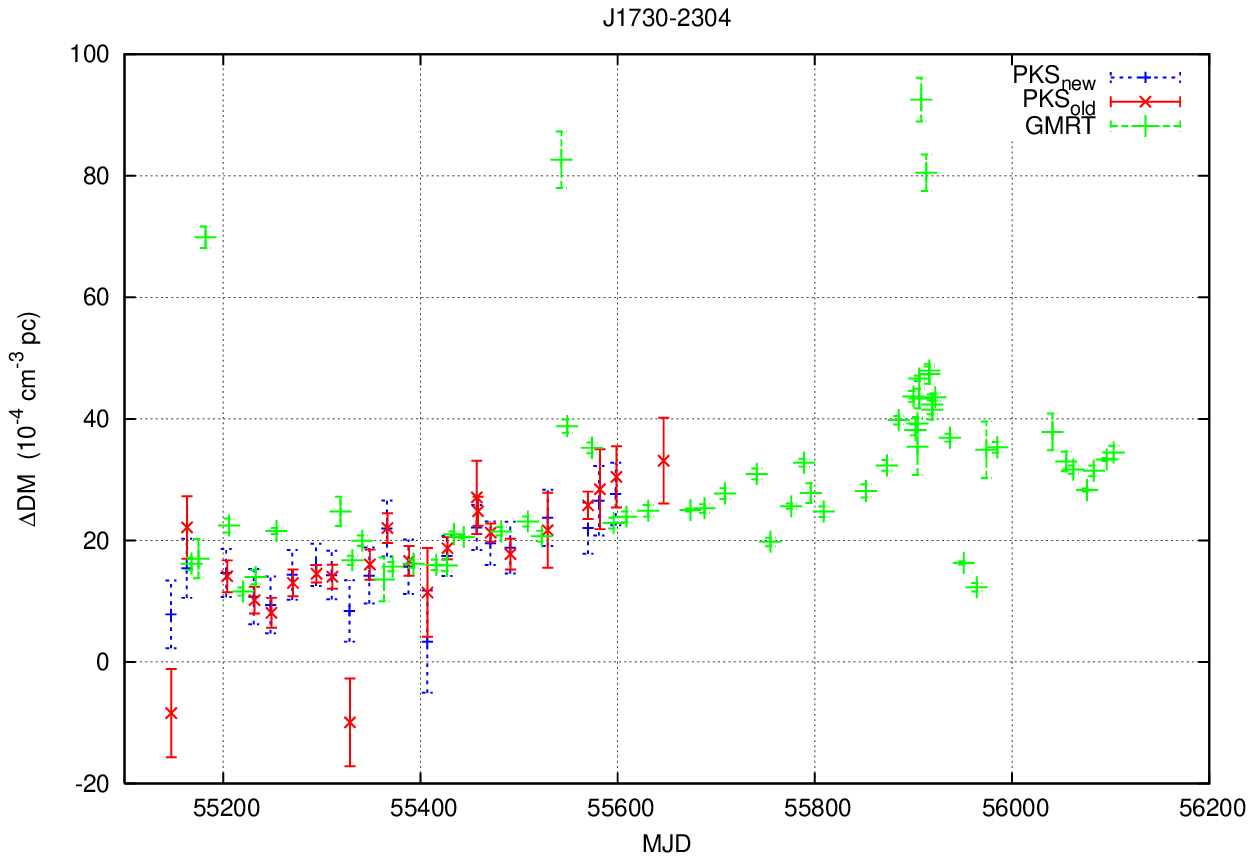}%
\includegraphics[height=2.0in,width=2.7in]{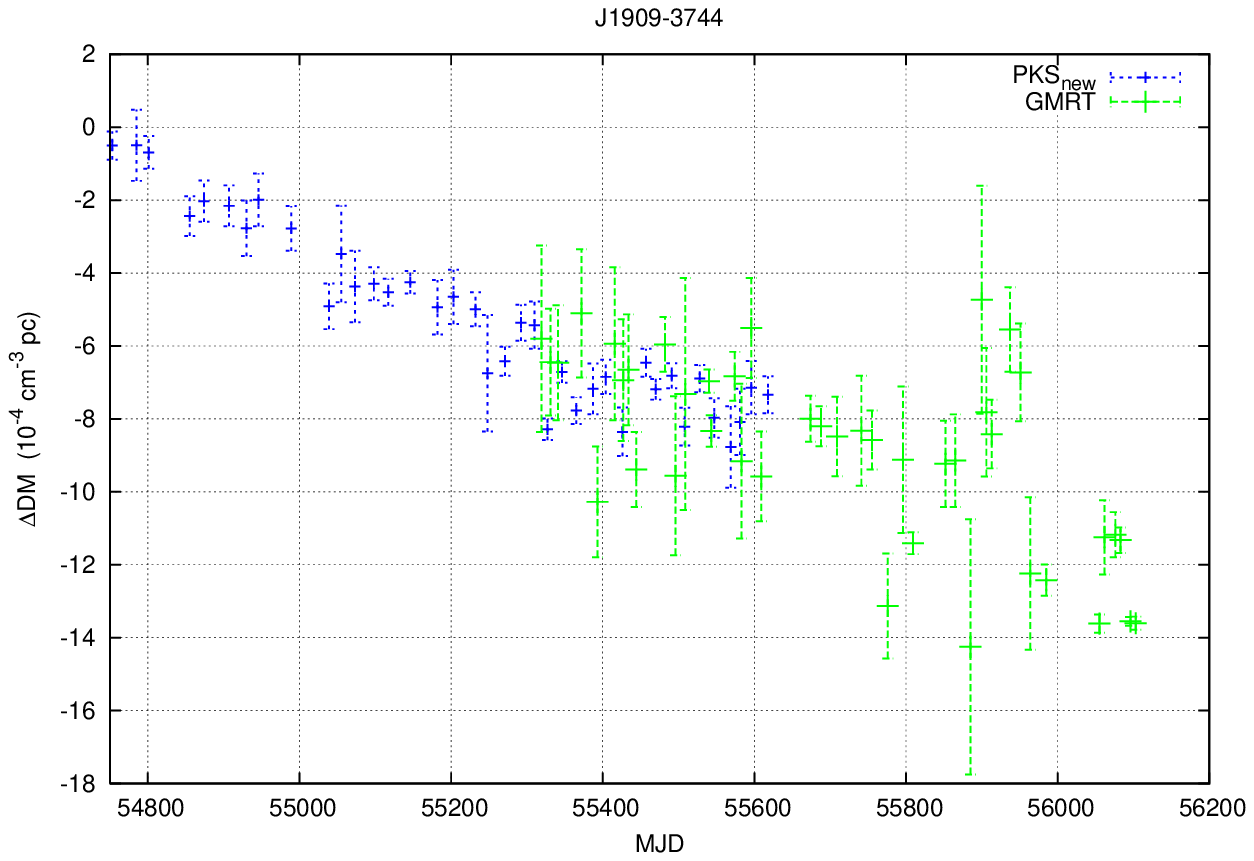} 
\caption{Comparison of DM time-series from the GMRT and Parkes, for PSRs J1909-3744 \& J1730-2304.}
\label{fig1}
\end{center}
\end{figure*}

Significant DM variations are detected (Table \ref{tab1}) for all the pulsars, with accuracies of $5 \times 10^{-4} \rm{~cm^{-3}~pc}$ achieved for most of them. 
For most pulsars, the rms DM variation is comparable to that seen in the PPTA data and also to the reported value from \cite{YouMN07}. The DM variations 
seem to show significant correlation with the Parkes data for four MSPs (Figure \ref{fig1}).

\begin{table}[tb]
\begin{center}
\caption{Summary of the DM measurements for nine of the MSPs. The catalogue period is in $\rm{ms}$, DM$_{cat}$ and $<DM>$ are in $\rm{~cm^{-3}~pc}$. 
rms$_{DM}$ and $<error>$ are in $10^{-4} \rm{~cm^{-3}~pc}$. The last four columns give the mean of DM, rms of DM, the mean absolute error of DM over 
all the epochs and the equivalent TOA error at L band (corresponding to the rms) in $\mu$s.}
\label{tab1}
\begin{tabular}{|c|c|c|c|c|c|c|}
\hline
{PSR}  & ${P_{cat}}$ & ${DM_{cat}}$ & ~${<DM>}$~   & ~${rms_{DM}}$~ & ~${<error>}$~   & ~${\Delta{TOA_l}}$~  \\
\hline
J2145-0750 & 16.0524       & 8.9977         & 9.0066      & 3.10             & 0.90              &       0.65                  \\
\hline
J1744-1134 & 4.07454       & 3.1390         & 3.1396      & 2.30             & 1.07              &       0.53                  \\
\hline
J1730-2304 & 8.12279       & 9.6170         & 9.6275      & 5.00             & 1.13              &       1.06                  \\
\hline
J1713+0747 & 4.57013       & 15.993         & 15.993      & 5.20             & 1.17              &       1.09                  \\
\hline
J1909-3744 & 2.94710       & 10.393         & 10.394      & 2.60             & 1.30              &       0.55                  \\
\hline
J1643-1224 & 4.62164       & 62.412         & 62.424      & 14.0             & 1.57              &       3.00                  \\
\hline
J0437-4715 & 5.75745       & 2.6447         & 2.6490      & 4.28             & ---               &       0.90                  \\
\hline
~J1022+1001~ & ~16.4529~       & 10.252         & 10.239      & 5.70             & 2.10              &       1.20                  \\
\hline
J0613-0200 & 3.06184       & ~38.779~         & ~38.795~      & 5.00             & 2.50              &       1.05                  \\
\hline
\end{tabular}
 \end{center}
\end{table}

\begin{figure*}[h]
\includegraphics[height=2.7in,width=2in,angle=-90]{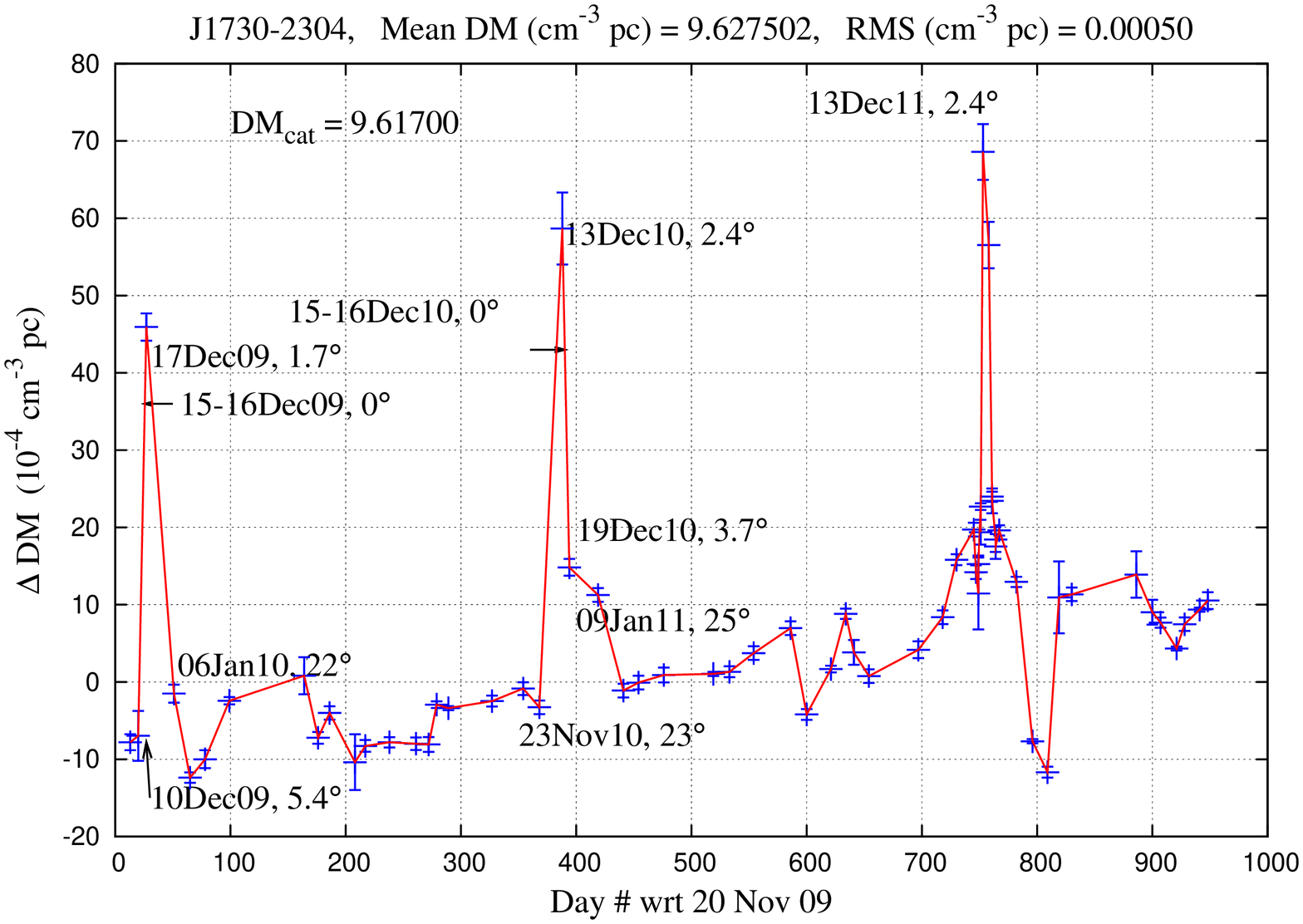} 
\includegraphics[trim=-28 0 0 0, clip, height=2.7in,width=2in,angle=-90]{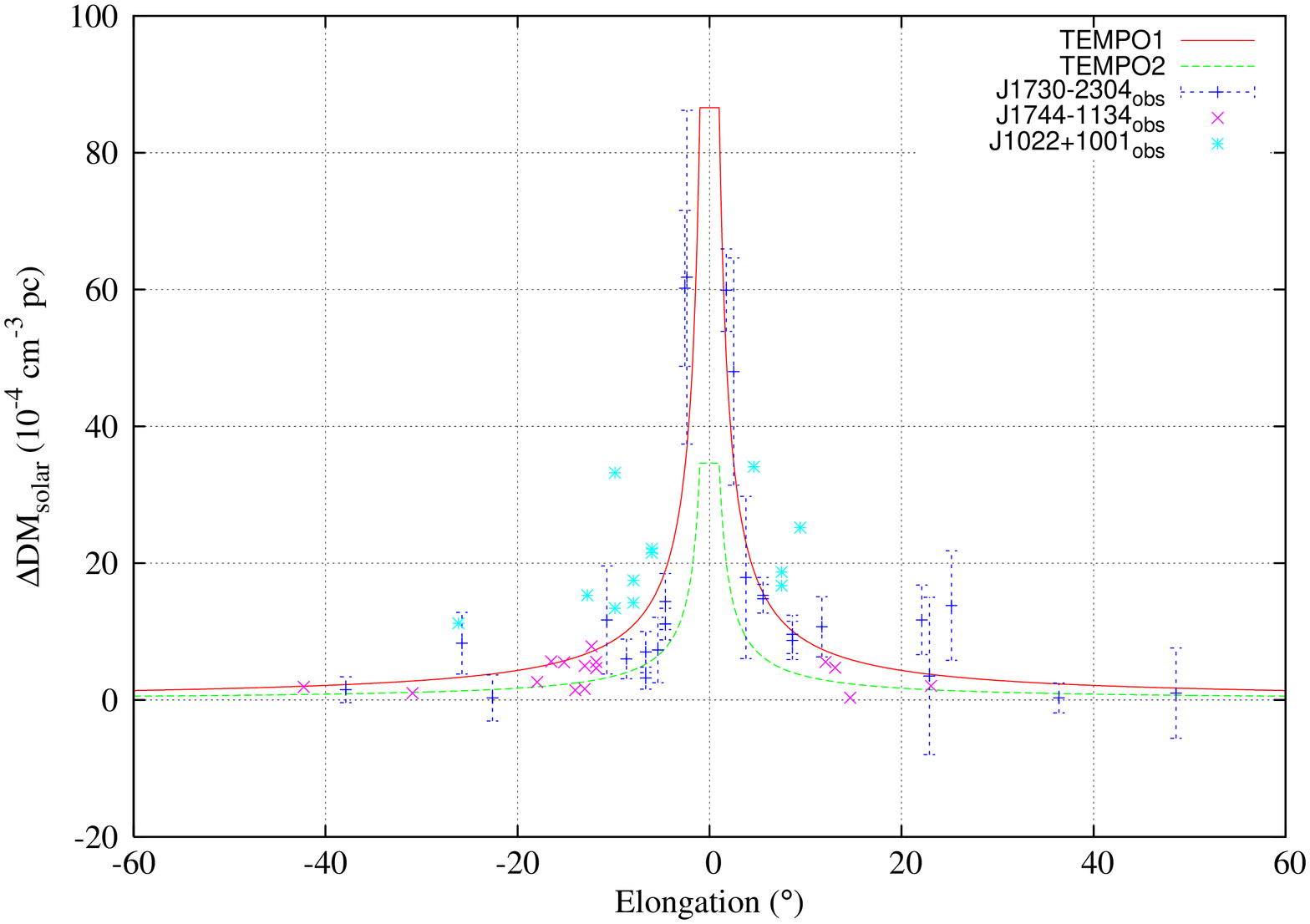} 
\caption{{\it Left:} DM variations for PSR J1730-2304, showing large increases around the times of closest approach to the Sun. 
{\it Right:} Consolidated DM variation as a function of elongation from the Sun from data for all pulsars and comparison with TEMPO1 (red) \& TEMPO2 (green).} 
\label{fig2}
\end{figure*}

The effects of solar corona are clearly detected in the case of low ecliptic latitude pulsars (Figure \ref{fig2}), even though many of our data points 
, up to 25$^{\circ}$ from the Sun, seem to disagree with the predictions based on the two-state solar wind model of \cite{YouAP07}, indicating possibilities
for further refinements of the model including effects of special events like coronal mass ejections (CMEs).

\begin{figure*}[h]
\includegraphics[height=5.4in,width=1.4in,angle=-90]{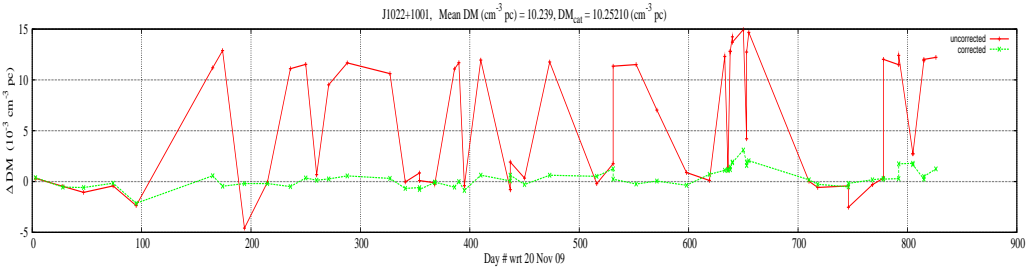} 
\caption{Two-state DM fluctuation for PSR J1022+1001.} 
\label{fig3}
\end{figure*}

Two-state jumps in PSR J1022+1001's DM variation (Figure \ref{fig3}) are found to be due to small, but quite well defined profile shape changes, akin 
to the well known mode changing phenomenon seen in some pulsars. Further comparisons and studies (e.g. ISM structure function analysis) will be possible 
in future as our data extend to longer time spans. 

\begin{thebibliography}{}

\bibitem[Ahuja \etal\ (2005)]{Ahuja05}
{Ahuja, A. L., Gupta, Y., Mitra, D., \& Kembhavi, A. K.} 2005,
\textit{MNRAS}, 357, 1013

\bibitem[Ahuja \etal\ 2007]{Ahuja07}
{Ahuja, A. L., Mitra, D., \& Gupta, Y.} 2007,
\textit{MNRAS}, 377, 677

\bibitem[Backer \etal\ (1993)]{Back93}
{Backer, D. C., Hama, S., \& Van Hook, S.} 1993,
\textit{ApJ}, 404, 636

\bibitem[Hobbs \etal\ (2004)]{Hobbs04}
{Hobbs, G. B., Lyne, A. G., Kramer, M., Martin, C. E., \& Jordan, C.} 2004,
\textit{MNRAS}, 353, 1311

\bibitem[Manchester 2011]{Manch11}
{Manchester, R. N.} 2011,
\textit{astro-ph.HE}, arXiv:1101.5202v1

\bibitem[Roy \etal\ 2010]{JRoy10}
{Roy, J., Gupta, Y., Pen, U. \etal\ } 2010,
\textit{astro-ph.IM}, arXiv:0910.1517

\bibitem[You \etal\ (2007a)]{YouMN07}
{You, X. P., Hobbs, G., Coles, W. A., Manchester, R. N. \etal\ } 2007,
\textit{MNRAS}, 378, 493

\bibitem[You \etal\ (2007b)]{YouAP07}
{You, X. P., Hobbs, G. B., Coles, W. A. \etal\ } 2007,
\textit{MNRAS}, 671, 907

\end{thebibliography}
\end{document}